\documentclass[12pt]{amsart}
\usepackage{epsfig,anysize,amssymb,verbatim}
\newsymbol\leqs 1336
\newsymbol\geqs 133E

\catcode`@=11
\def\@oddfoot{\hfil \thepage \hfil}
\def\@evenfoot{\hfil \thepage \hfil}
\def\@oddhead{\hfil}
\def\@evenhead{\hfil}
\catcode`@=12

\begin{document}
\large
\centerline{\large\bf Synchrotron radiation in}
\centerline{\large\bf cyclic accelerators}

\vskip3mm
\centerline{{O.E.Shishanin}\footnote{e-mail: olegsh55@gmail.com}} 

\vskip3mm

\begin{center}
                    Moscow State University of Food Production\\

                      Department of Physical and Mathematical Sciences\\
                          125080 Moscow, Volokolamskoe highway, 11, Russian Federation \\
                          
\end{center}

\bigskip

Abstract
\begin{center}

{\small
 A techniques, describing electron dynamics for magnetic models closed to
  cyclical accelerators is
developed and applied to the analysis of electromagnetic radiation
emitted by charged particles.
 Formulas for the angular characteristics of synchrotron light, which
take into account the electron vibrations in the lattices
of accelerators and storage rings, are derived.
 It is shown that both the degree  of photon polarization, as well as
the spectral and angular  distributions of radiation intensity, exhibit an observable
 dependence on the vertical betatron oscillations.}
\end{center}

\vskip3mm


\vskip5mm


\bigskip

1. Introduction

\vskip5mm

 At present there is a continuing expansion in the construction of
synchrotron radiation sources and their utilization in various fields
 of physics and technology. In view of this, the development of more
  accurate methods for studying
the radiation characteristics of these unique experimental facilities
is warranted.
 The existing theory of  synchrotron radiation has heretofore
 been developed mainly for
uniform  magnetic fields ~\cite{Sch1,Sch2,Sok1,Sok2}. As a consequence, the formulas
 obtained are not generally valid  in the field lattices of modern
accelerators and storage rings.   The
agreement of conventional theories  with experimental data
is satisfactory mainly for the parameters of 
 spectral density and  total power.
 In actuality,  electrons  accelerating in periodic magnetic 
 fields  perform transverse 
oscillations which can noticeably 
affect the angular and polarization properties of the emitted radiation.
In consequence of this, it would appear advisable elaboration to formulate the 
emission problem from first to last 
 in a non-uniform magnetic field. Such an analysis would, among other things,
motivate the development of more comprehensive
beam  diagnostics  and enhance the precision of experiments involving
polarizable synchrotron radiation.

  The effect of betatron oscillations on synchrotron
radiation was first discussed 
 in Ref. ~\cite{Zhu} for an axisymmetric magnetic
field.  Then, we carried out an analysis 
of  the properties of radiation emitted in the FD,
FODO, FOFDOD systems, etc. ~\cite{Sh1,Sh2,Sh3}.  In the present
 paper, using the beta function concept,
 we attempt to extend the theory to more general magnetic structures,
 including storage rings.

\vskip5mm

2. Betatron function method

\vskip5mm
  According to present views ~\cite{Bru},  the small vertical motions of charged particle in a
circular accelerator can be described by the function
$$
z=\sqrt{\frac{\beta_z A_z}{\pi}}\cos(\int \frac{ds}{\beta_z}+
  \delta_0), \eqno (1)
$$
where $A_z$ is the emittance, $\beta_z$ is the beta function that 
depends on the orbital
length $s$, and $\delta_0$ is the initial phase.

  Let $\varphi=s/R_0$ be a generalized azimuthal angle, where
  $R_0$ is the mean radius (viz., the total orbit length  
 devided by $2 \pi$ ) in magnetic systems with straight sections.

  In this case the velocity component is determined from (1) to be
$$
v_z=c \sqrt{\frac{A_z}{\pi \beta_z}}\sqrt{1+(\frac{1}{2}\frac{d
\beta_z}{ds})^2} \cos(\int\frac{ds}{\beta_z}+\delta_1+
\delta_0), \eqno (2)
$$
  where $\sin\delta_1 = 1/ \sqrt{1 + (d \beta_z/ds)^2/4}.$

In cylindrical co-ordinates, if the particle position is specified
 by the radius-vector
$$
 \{(R_0 +\rho) \cos\varphi,\, (R_0+\rho)\sin\varphi,\, z\},
$$
the velocity components are given by
$$
v_x =\dot \rho \cos\varphi-(R_0+\rho)\sin\varphi \,\dot \varphi, 
v_y = \dot \rho\sin
\varphi + (R_0 + \rho) \cos\varphi\, \dot \varphi,v_z =\dot z,
$$
where the dot denotes the total derivative with respect to t.

When $\rho$ is added to $R_0$, the
direction of a narrow cone of radiation varies 
moderately.
Thus the radial oscillations influence the angular radiation 
distributions only slightly and  we may assume that
 the particle moves on an average circular orbit of radius $R_0$.

To obtain a continuous description of particle motion,
the guiding magnetic field will be extended over the entire 
 length of the lattice. In this case the field can be written as
$$
H_0(\frac{1}{1+k} +F(\varphi)), \eqno (3)
$$
where $k$ is the ratio of lengths of the straight sections to the
lengths of the bending magnets, and the
 periodic function $F(\varphi)$  averages out to zero.
Following a Fourier series expansion,
 the transverse projections of the magnetic field
take the form:
$$
H_z=H_0(\frac{1}{1+k}-f(\varphi)\frac{\rho}{R}),\quad 
H_r=-H_0 f(\varphi)\frac{z}{R}, \eqno (4)
$$
where $R$ is the usual magnet bending radius and  
 the step function $f(\varphi)$, introduced in place of the gradient
 distribution $n$,
has a definite form associated with any given magnetic structure.

 In order to get an expression for the angular velocity
 one can take as a basis the following equation   
$$
\frac{1}{r}\frac{d}{dt}(mr^2\dot \varphi)=-\frac{e_0}{c}
(\dot z H_r-\dot r H_z), \eqno (5)
$$
where the terms $H_r$ and $H_z$ are generally different 
 for each element of a cell.
 
After substituting (4) into (5) and integrating we have,
in the parabolic approximation
$$
r^2 \dot \varphi=\omega_0[R R_0+R \rho +\frac{1}{2(1+k)} \rho^2+
(1+k)\int f(\varphi)(z\dot z-\rho\dot \rho)dt],
$$
where $\omega_0=ce_0H_0/E,\,\,E$ is the total electron energy,
 and $R_0=(1+k)R$.
Integrating here $\dot r$ restored former state $r$ as $R_0 +\rho$.
As a result of  new value of frequency
take into account also straight sections.

  The final expression for the angular velocity assumes the form: 

\includegraphics{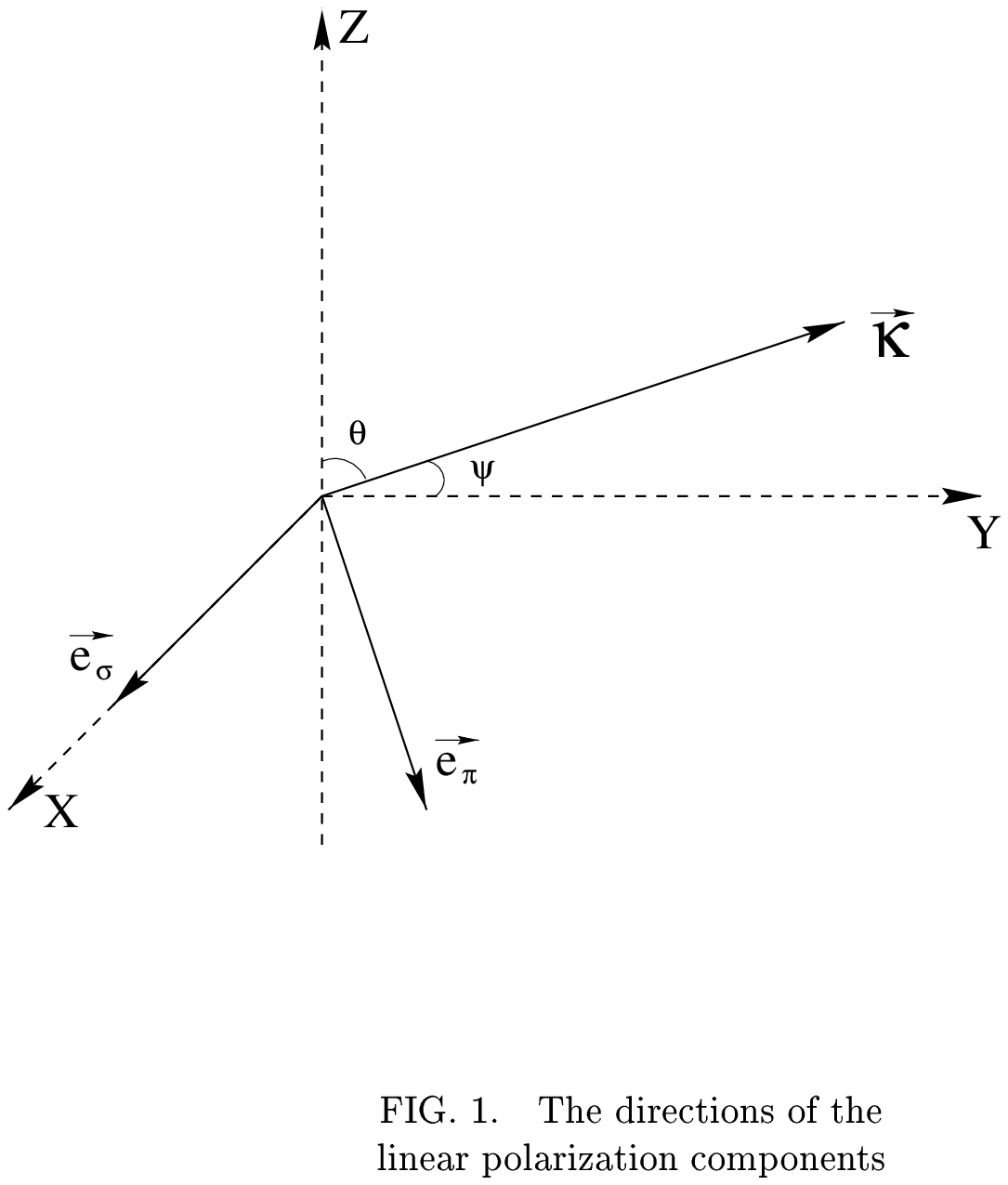}

$$
\dot \varphi=\frac{\omega_0}{1+k}[1-\frac{\rho}{R_0}+\frac{3}{2}
\frac{\rho^2}{R_0^2}+\frac{1}{R^2}\int(z\dot z-\rho\dot \rho)f(\varphi)
dt]. \eqno (6)
$$
  
  We shall now employ the operator method ~\cite{Sch2,Zhu} for studying 
the properties  of
the emitted synchrotron light. We will
suppose that the radiation vector $\vec \varkappa
=\omega \vec n/c$, where
$\vec n=\{0;\sin\theta,\cos\theta\}$
lies in the YZ plane (see also ~\cite{Lan}),
 and $\theta$ is the spherical angle.

  We define the linear polarization vectors in the wave zone as follows:
$$
\vec e_{\sigma}=\{1,0,0\},\qquad \vec e_{\pi}=\{0,
\cos\theta,-\sin\theta\}.
$$

  Here, the first vector corresponds to the
 electric field component lying in the
orbital plane and the second is orthogonal to this plane (see Fig.1).
 In the first quantum approximation the components of the emission
intensity can be written as
$$
\frac{dW_{\sigma}}{d^3\varkappa}=
\frac{ce^2}{(2\pi)^3R_0}\frac{\nu'}{\nu}\mid\int
 dt\, v_x \exp(i \frac{\nu'}{\nu}(\omega t-\vec \varkappa \vec r))\mid^2,
$$
$$
\eqno (7)
$$
$$
\frac{dW_{\pi}}{d^3\varkappa}=\frac{ce^2}{(2\pi)^3R_0}\frac{\nu'}{\nu}\mid\int
 dt( v_y \cos\theta-v_z \sin\theta) 
\exp(i \frac{\nu'}{\nu}(\omega t-\vec \varkappa \vec r))\mid^2,  
$$
\medskip
where $\omega=\nu \omega_0/(1+k)$ and $\nu' =\nu(1 + h\,\omega/E).$

  The  angle $\varphi$  can  be  reckoned  from  any  point  of  the
  orbit.   Then,
the  expansion  parameters  are given by  $\varphi \sim m_0 c^2/E,
\tau=N\varphi$ ($N$  is the
number of  magnetic periods in the accelerator or storage ring),
$\cos\theta(\theta\sim \pi/2),\rho/R_0$, and $z/R_0.$
 Using these, we also take $\sin \varphi\sim\varphi-\varphi^3/6,
\sin\theta\sim1$ in the parameter
$$
\Phi=\omega t-\vec \varkappa \vec r=\frac{\nu\omega_0}{1+k}[t-
\frac{1}{c}(R_0+\rho)\sin\varphi\sin\theta-\frac{z}{c}\cos\theta].
$$

  To determine  $\varphi$ in (6) it is necessary
 to carry out the integration. In particular, in
the zeroth approximation $\varphi =\omega_0 t/(1 + k).$

  From the equality
$$
v^2=\dot \rho^2+\dot z^2+R^2\omega_0^2[1+\frac{\rho^2}{R_0^2}+
\frac{2}{R^2}\int(z\dot z-\rho\dot\rho)f(\varphi) dt]
$$
we can solve for the quantity $R  \omega_0/c$.

  The velocity components are
$$
\frac{v_x}{c}\approx \frac{\dot \rho}{c}-\varphi, \quad
\frac{v_y}{c}\approx\beta,\quad  v_z=\dot z.
$$

  We now introduce a new variable $u = -v_x/c=\varphi-\dot \rho/c$
and carry out expansions of the form:

$$
z=z\mid_{\tau=0}+\frac{dz}{d\tau}\mid_{\tau=0}\cdot\tau+...
\approx z_0+\frac{v_z}{c}
R_0 \varphi,
$$
$$
\frac{\omega_0}{1+k}\int dt [\int(z\dot z-\rho \dot \rho) f(\varphi)dt]
\approx const+\frac{1}{N}\tau\int dt (z\dot z-\rho \dot\rho)f(\varphi)
$$
and so on.

  Retaining  terms up to the third order, we obtain
$$
\Phi=[1-\beta\sin\theta+\frac{u^2}{6}-\frac{v_z}{c}\cos\theta+
\frac{1}{2}(\frac{v_z}{c})^2] u+const.
$$

  In the ultrarelativistic case we have $1 - \beta\sin\theta \approx
\varepsilon/2$    , where $\varepsilon=1-\beta^2\sin^2\theta.$

  Integrating Eqs.(7) with respect to the new variable $u$  and averaging
over the initial phases we get for the spectral-angular distributions

$$
\frac{dW_{\sigma}(\nu)}{d\Omega}=\frac{ce^2\nu\nu'}{12\pi^4 R_0^2}
\int_0^{2\pi} d\delta\,\varepsilon_1^2\,K_{2/3}^2(\frac{\nu'}{3}
\varepsilon_1^{3/2}),
$$
$$
\eqno (8)
$$

$$
\frac{dW_{\pi}(\nu)}{d\Omega}=\frac{ce^2\nu\nu'}{12\pi^4 R_0^2}
\int_0^{2\pi} d\delta\,\varepsilon_1\,\varepsilon_2\, K_{1/3}^2(\frac{\nu'}{3}
\varepsilon_1^{3/2}),
$$
where
$$
\varepsilon_1=1-\beta^2+\varepsilon_2,\qquad \varepsilon_2=(\cos\theta-
\frac{v_z}{c}\mid_{\tau=0})^2,
$$
$$
\frac{v_z}{c}\mid_{\tau=0}=\alpha\,\cos\delta, \qquad \alpha=\sqrt{
\frac{A_z}{\pi}}\left[\frac{1}{\sqrt{\beta_z}}\sqrt{1+(\frac{1}{2}\frac
{d\beta_z}{ds})^2}\right]\mid_{\tau=0}.
$$

  In (2) in the neighbourhood of the point $\tau =0$ the constant phase was
denoted by $\delta$ .

  Formulas (8) can mainly be used  for storage rings.  For
 their practical application
 we must bear in mind that the graph of the beta function is usually
known for a closed trajectory. Consequently, we can define
 its value at the point at
which emission is recorded, while the derivative can be approximated by
the ratio $\bigtriangleup\beta_z/\bigtriangleup s$.

  For bending magnets the graph of the beta function is almost linear;
in this case we can use  the slope of the linear segment for approximating
 the derivative.

Introducing  the angle $\psi$ reckoned from the orbital plane,
we can approximate  $\cos\theta$ by $\psi$.

Let us now discuss  the method of calculating  the integrals in (8).
Initially, we may go over to the Airy function
$$
K_{1/3}(\frac{2}{3}x^{3/2})=\pi\sqrt{\frac{3}{x}} Ai(x), \qquad
K_{2/3}(\frac{2}{3}x^{3/2})=-\frac{\pi\sqrt{3}}{x} Ai'(x)
$$
with $x=(\nu/2)^{2/3}\,\varepsilon$ and employ well-known
 integral tables ~\cite{Abr}.

 In the case of greatest interest (for small amplitudes
of the oscillations or small cross-sections of the bunch) one can use an
expansion  in  terms of the parameter $q^2=\alpha^2/2\varepsilon$.
Then, in the classical approximation we obtain the following
expressions:
$$
\frac{dW_{\sigma}(\nu)}{d\Omega}=W\{(Ai')^2+2x^2q^2[2xgU+(1+2g)AiAi']+
\frac{1}{2}x^2q^4[12x^2g(1+g)Ai^2+
$$
$$
x(3+24g+16x^3g^2)U+3(1+16x^3g+24x^3g^2)AiAi']\},
$$
$$
\eqno (9)
$$
$$
\frac{dW_{\pi}(\nu)}{d\Omega}=Wx\{(g+q^2)Ai^2+2xgq^2(2xgU+5AiAi')+
3xq^4 AiAi'+
$$
$$
\frac{1}{2}x^2gq^4[(39+16x^3g^2)U+28xgAi^2+8x^2g(14+3g)AiAi']\},
$$

where
$$
W=\frac{2^{1/3}ce^2\nu^{2/3}}{\pi^2 R_0^2},\qquad g=\psi^2/\varepsilon,
\qquad U=xAi^2+(Ai')^2,
$$
$$
 \varepsilon=1-\beta^2+
\cos^2\theta=\frac{1}{\gamma^2}(1+\gamma^2\psi^2).
$$
We note that the parameter $x$ is the argument both of the function $Ai$
and its derivative. 

Assuming  $q^2=0$ (there are no vertical oscillations)
 and replacing $R_0$
by $R$, these formulas can be transformed into expressions for 
an uniform  magnetic field.

In the neighbourhood of the orbital plane  $x$ is  the suitable expansion
parameter. Here we use the Airy function 
$V(x)=\sqrt{\pi}Ai(x)$ with the initial Fock conditions:
$V(0)=0.629271, V'(0)=-0.458745$ ~\cite{Jak}.                       
This function, along  with its derivative, can be expanded in 
convergent power series. In this case, the braces in (9)
must be replaced by, respectively,

$$
V'(0)^2\{1+\frac{1}{3}x^3[2+3q^2(2+3q^2)(1+4g)+5q^6]\}+
$$
$$
V(0)V'(0)x^2(1+2q^2+4q^2g+\frac{3}{2}q^4)
$$
and
$$
V^2(0)(g+q^2)+V(0)V'(0)x(2q^2+3q^4+2g+10gq^2)+
$$
$$
\frac{1}{2}V'(0)^2 x^2[2q^2+6q^4+5q^6+g(2+20q^2+39q^4)+8q^2g^2].
$$

 From these formulas we can observe that  at $\theta=\pi/2$ the 
$\sigma$-component is less than the same component 
 for a uniform magnetic field,
and the magnitude of the $\pi$-component is not equal to zero.

 Peaks of curves plotted in accordance with formulas (8) will generally be
  below the graphs that are built to 
 the uniform  magnetic field at the same energy $E$ and bending radius $R$.
In the plane of the equilibrium  orbit 
 the radiation will not be  completely
linearly polarized.

\vskip5mm

3. Angular characteristics of synchrotron light 

\vskip5mm

 We will first discuss the spectral properties of the emission.
If, in (9), we retain  $\cos\theta$ instead of $\psi$
 and integrate with respect to the 
spherical angle $\theta$, we derive  spectral formulas almost coinciding
with the analogous expressions for a uniform  magnetic field.
 There are corrections to the order of  
$\alpha^2/R_0^2$ , and  similar terms are  to the radial vibrations.

Summing expressions (8) over the entire spectrum, we come
to the following integrals:
$$
\frac{dW_{\sigma}}{d\Omega}=\frac{7ce^2}{64\pi^2 R_0^2}\int_0^{2\pi}
d\delta\,(\frac{1}{\varepsilon_1^{5/2}}-\frac{320}{7\sqrt{3}\pi}
\frac{1}{\varepsilon_1^4}\frac{h\omega_0}{E}), \eqno (10)
$$
$$
\frac{dW_{\pi}}{d\Omega}=\frac{5ce^2}{64\pi^2 R_0^2}\int_0^{2\pi}d \delta
\,(\cos\theta-\alpha\cos\delta)^2\,(\frac{1}{\varepsilon_1^{7/2}}-
\frac{256}{5\sqrt{3}\pi}\frac{1}{\varepsilon_1^5}\frac{h\omega_0}{E}).
$$

If we now integrate with respect to 
 $\theta$ and $\delta$, we obtain the total
intensities:

$$
W_{\sigma}=\frac{7ce^2}{12 R_0^2\varepsilon_0^2}(1-\frac{25\sqrt{3}}
{7\varepsilon_0^{3/2}}\frac{h\omega_0}{E}),\qquad
W_{\pi}=\frac{ce^2}{12 R_0^2\varepsilon_0^2}(1-\frac{5\sqrt{3}}
{2\varepsilon_0^{3/2}}\frac{h\omega_0}{E}).
$$
where  $\varepsilon_0=1-\beta^2$.
Substituting  $R_0$ by $R$ and summing these components we come to
 the well-known result ~\cite{Sch2,Sok1}.

Methods of integrating the first terms in (10)  
with respect to  $\delta$
were considered in ~\cite{Zhu}.
For the  final calculations  it is convenient to
introduce the additional variables:
$$
\varepsilon=\varepsilon_0+\cos^2\theta,\quad p=\alpha^2/\varepsilon,
\quad g=\cos^2\theta/\varepsilon,\quad p_1=\alpha^2/\varepsilon_0,
\quad g_1=\cos^2\theta/\varepsilon_0,
$$
$$
f=\varepsilon_0/\varepsilon,\quad  \triangle=(1+p)^2-4pg,\quad
2r^2=1-(1-p)/\sqrt{\triangle}.
$$ 

First, we need to express $\int_0^{2\pi}d\delta/\sqrt{\varepsilon_1}$
through the complete elliptic integral ${\bf K}(r)$, then to differetiate
 this eguality several times in the partameters; here it is necessary 
 to use the equation
 $$
 x(1-x)\frac{d^2{\bf K}}{dx^2}+(1-2x)\frac{d{\bf K}}{dx}-\frac{1}{4}{\bf K}=0,
 $$
 taken from ~\cite{Jahn}.

In the ultrarelativistic case
the classical part of the angular distributions of the radiation
intensity  can be written in the form:

$$
\frac{dW_{\sigma}}{d\Omega}=\frac{14 W_1}{3\pi\varepsilon^{5/2}
\triangle^{5/4}}\{(3+p_1+16\frac{pg}{\triangle}-\frac{2p}
{\triangle^{1/2}r^2}G_1){\bf K}+\frac{2\triangle^{1/2}}{f}G_1
{\bf E}\},
$$
$$
\eqno (11)
$$
$$
\frac{dW_{\pi}}{d\Omega}=\frac{2W_1}{3\pi\varepsilon^{5/2}
\triangle^{5/4}}\{(G_2-\frac{p}{\triangle^{1/2}r^2}G_3)
{\bf K}+\frac{\triangle^{1/2}}{f}G_3{\bf E}\},
$$

\medskip
where {\bf K}(r) and other elliptic integral {\bf E}(r) calculated
using tables ~\cite{Jahn}, and
$$
W_1=ce^2/32\pi R_0^2,\qquad G_1=p_1-g_1+(2/\triangle)[(p+f)^2-g^2],
$$
$$ 
G_2=p_1+\frac{1}{\triangle}[8p(p+f)-25pg+15g]+\frac{8pg}{\triangle^2}
[9(p-g)^2-7f^2+2f(p+g)],
$$
$$
G_3=1+2(p_1-g_1)+\frac{1}{\triangle}[4(p^2-f^2)+3g(p-7f)-7g^2]+
$$
$$8\frac{gf}{\triangle^2}[7-9p^2+2p(g-f)].
$$

If the betatron oscillations are small $(p<<1)$ we can develop
Eqs.(11) in powers of $p$. In this case we get the more obvious 
formulas:
$$
\frac{dW_{\sigma}}{d\Omega}=\frac{7W_1}{\varepsilon^{5/2}}
(1-\frac{5}{4}p+\frac{35}{4}pg),
$$
\medskip
$$
\frac{dW_{\pi}}{d\Omega}=\frac{5W_1}
{\varepsilon^{5/2}}(g+\frac{1}{2}p-
\frac{35}{4}pg+\frac{63}{4}pg^2).
$$

Setting  $p=0$ and replacing  parameter $R_0$ with $R$ in $W_1$
 we regain  the usual expressions for
a uniform magnetic field.

This section of the most well  applicable near the 
 critical wavelength of the synchrotron light.

\vskip5mm

4. Perturbation method

\vskip5mm
In additional,  we would like to attach to the  given topic the magnetic systems of cyclic accelerators.
As an illustration we shall exemplify a case
 based on 
 averaging   dynamic characteristics of the strong-focusing
FODO structure. At this point, we will
 denote the lengths of bending
magnets and straight sections 
by $a$  and $l$,  respectively.

Let $N$ be the number of periods and $L=2a+2l$ be the length of a
single cell. The length of closed trajectory is equal to
$2\pi R+2Nl=2\pi R_0$, where $R$ is the radius of magnetic
curvature and  $R_0$ is the so-called mean radius.
With this,  we again obtain
  $R_0=(1+k)R$, where $k=Nl/\pi R=l/a.$

Since we are interested primarily in small betatron oscillations, the 
 same simplifications concerning the parameters of the
main orbital motion can be used again.
We suppose  that a particle performs a rotation of radius $R_0$
in an average guiding magnetic field $H_0$ over one period.
Note that in this case the periodic function $F(\varphi)$ entering in (3)
is equal to
$$
\frac{2}{\pi}\sum_{\nu=1}^\infty\frac{\sin2\nu\tau_1}{\nu}
\cos2\nu(\tau-\tau_1),
$$
where $\tau=N\varphi,\,\,\tau_1=\pi a/2(a+l).$

The components of the magnetic field now take the form:
$$
H_z=H_0(\frac{a}{a+l}-f(\tau)\frac{\rho}{R}),\qquad
H_r=-H_0f(\tau)\frac{z}{R},
$$
where $\rho=r-R_0$,
$$
f(\tau)=\frac{4n}{\pi}\sum_{\nu=0}^\infty\frac{1}{2\nu+1}
\sin(2\nu+1)\tau_1\,\cos((2\nu+1)(\tau-\tau_1)).
$$

In the linear approximation
the betatron oscillations of a charged particle may be described
 by the following equations:
$$
\frac{d^2\rho}{d\tau^2}+\frac{1}{N^2}(1-(1+k)^2f(\tau))\rho
=0,
\eqno (12)
$$
$$
\frac{d^2z}{d\tau^2}+\frac{1}{N^2}(1+k)^2f(\tau)z=0.
\eqno (13)
$$

To complete a description of the full system we must add  Eq. (6). 
According to the  Floquet theorem a solution 
 for $\rho$ or $z$ may be expressed in the form
 $\exp(i\gamma\tau)\,\varphi(\tau)$ provided that $\varphi(\tau+
2\pi)=\varphi(\tau).$

Replacing $\rho$ and $z$ in (12) and (13),
 we obtain  new differential equations for the
periodic functions $\varphi(\tau).$ We choose $1/N$ as the expansion
parameter and develop $\varphi(\tau)$ and $\gamma$ in power series.
Equating coefficients of like powers to zero, we come to a chain
of differential equations. As part of our procedure we also have
to eliminate  the secular terms. Finally, we obtain
the following asymptotic forms:
$$
\rho=A\cos(\frac{\nu_\rho}{N}\tau+\chi_0)(1-S_1)-A\sin(\frac{\nu_\rho}
{N}\tau+\chi_0)S_2,
\eqno (14),
$$
$$
z= B\cos(\frac{\nu_z}{N}\tau+\delta_0)(1+S_1)+B \nu_z \sin(\frac{\nu_z}{N} 
\tau +\delta_0)S_2, \eqno (15)
$$

where
$$
S_1=G\sum_{\nu=0}^\infty q_{\nu} \cos(2\nu+1)
(\tau-\tau_1),\qquad g_{\nu}=\frac{\sin(2\nu+1)\tau_1}{(2\nu+1)^3},
$$
$$
S_2=\frac{2G}{N}\sum_{\nu=o}^\infty\frac{g_\nu}{2\nu+1}
\sin(2\nu+1)(\tau-\tau_1),\qquad \nu_z=\frac{\pi n}{2\sqrt{3} N}
\sqrt{1+4k+3k^2},
$$
$$
G=4n(1+k)^2/\pi N^2,\qquad \nu_\rho=\sqrt{1+\nu_z^2}.
$$

Here,
$A$ and $B$ are the amplitudes of the dominant sinusoidal motions, and
$\chi_0$ and $\delta_0$ are the initial phases. Moreover, the frequency
$\nu_z$ is the same as shown in ~\cite{Bru} for this structure.

Eq. (15)  can now be written in the form:
$$
z=B \sqrt{1+2S_1}\cos(\frac{\nu_z}{N}\tau+\delta_1+\delta_0)  \eqno (16)
$$
with $\sin\delta_1 \approx \nu_z S_2$.

Comparing Eq.(1)  with Eq.(16), we can see that $\beta_z A_z/\pi=
B^2(1+2S_1).$ After averaging, we have $\bar \beta_z A_z/\pi=B^2.$
Bearing in mind that $\bar \beta_z=R_0/\nu_z$, we obtain the
 approximation  $B\approx \sqrt{A_z R_0/\pi \nu_z}.$

Thus, we have $\beta_z=R_0(1+2S_1)/\nu_z$ and
$$
\frac{d\beta_z}{ds}=-\frac{8n(1+k)^2}{\pi N\nu_z}\sum_{\nu=0}^\infty
g_\nu\, (2\nu+1)\,\sin(2\nu+1)(\tau-\tau_1).
$$

Using Eq.(2), the parameter $\alpha$ entering into (8) can, finally, be
determined by the substitution  of  values

$$
\alpha= \sqrt{\frac{A_z\nu_z}{\pi R_0}} \cdot
 \sqrt{1+\frac{\pi^2n^2(1+k)^2}
{4 N^2\nu_z^2}-\frac{\pi^2 nk}{2 N^2}}.
$$

This makes it possible to investigate the angular properties of 
the emitted radiation.

\includegraphics{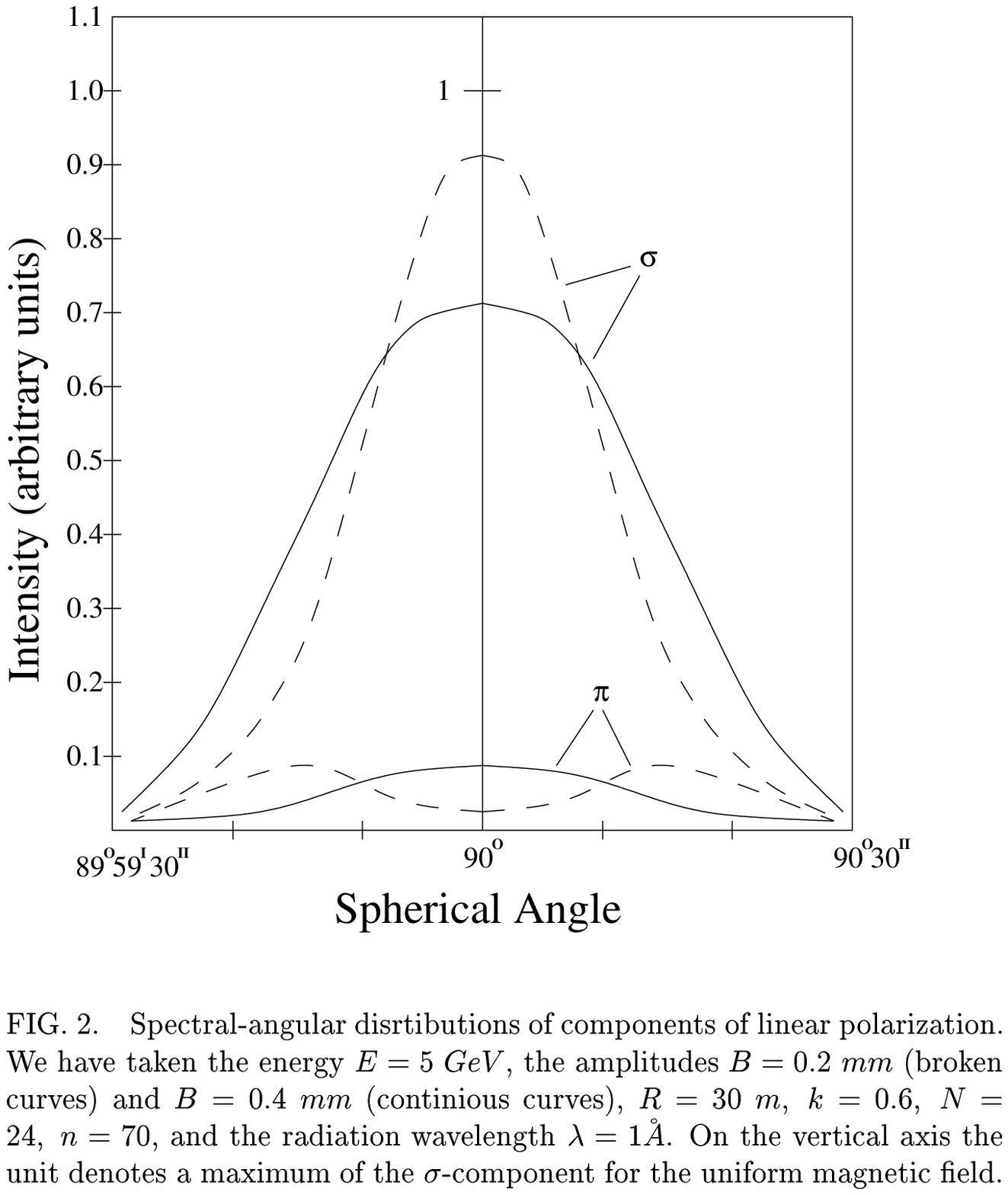}

\newpage

V. Comparison with experiments

\vskip5mm

An example of spectral-angular distributions of radiation
 in the FODO lattice is shown in Fig.2.
Here the angle $\theta=90^\circ$ corresponds to the orbital plane.
As required we can easily recount degrees in radians
on the horizontal axis.
The indicated  plots  demonstrate the effect of  betatron oscillations
on the angular radiation properties.

Special experiments were carried out on this issui for a long time on
the electron synchrotrons with different energies ~\cite{Sok1,Vor}.
In general, we can note a good agreement  with the theory of these works.
All authors emphasize the important role of the axial electron oscillations 
in the formation of the spectral-angular properties.

Basic formulas (8) determine the properties of the radiation for modern accelerators.
Here the parameters $\cos\theta$ or $\psi$, which is often used in literature, add
in $\varepsilon_2$ a scalar term, formed by variable vector of particle velocity.  
Thereby the radiative scattering increases.

That can give these expressions at large amplitudes of oscillations?
 At  higher oscillation amplitudes  the 
$\pi$-component exhibits a peak instead of the minimum at $\theta=\pi/2.$
Furthermore, it is expected that the 
$\sigma$-component will attain  a small local minimum in the orbital plane
for the extremal amplitudes
and, in addition,  the $\pi$-component will generate symmetric local
hollows.

Thus, based on our theoretical study  we propose
to search  these effects in
 experiments on existing machines  with a higher degree of accuracy 
than previously.
 We are talking only about the storage rings. Moreover, these tests were
 at fixed wavelengths and, apparently, experiments should be carried out also for
 the total angular distributions. 
 Because  the radiation, that  we  consider, is incoherent at 
wavelengths which cause the  greatest practical interest our  results
can be applied to the electron beams.
Specifically, the cross-section of a typical
bunch from which the radiation is emitted constitutes
 an  ensemble of charged particles with 
 various amplitudes. Recognizing this, we can re-interpret the 
amplitude parameter $B$  in our formulas as a mean square value
using following arguments.
We take into account the longitudinal beam distribution, specify  the injector
scatter of electrons and carry out averaging over initial phases.

\end{document}